\documentclass[floatfix,nofootinbib,superscriptaddress,twocolumn,pre,showpacs,showkeys,preprintnumbers]{revtex4-1}
\usepackage{amsmath,amssymb}
\usepackage{epsfig,graphics,color,calc,graphicx}
\usepackage{mathrsfs}

\newcommand{\rr}[1]{{\normalfont\textrm{#1}}}

\newcommand{\bb}[1]{{\mathbb{#1}}}

\newlength{\pecettawidth}
\begin{document}
\title{Compacton formation under Allen--Cahn dynamics}

\author{Emilio N.M.\ Cirillo}
\email{emilio.cirillo@uniroma1.it}
\author{Nicoletta Ianiro}
\email{nicoletta.ianiro@uniroma1.it}
\affiliation{Dipartimento di Scienze di Base e Applicate per l'Ingegneria, 
             Sapienza Universit\`a di Roma, 
             via A.\ Scarpa 16, I--00161, Roma, Italy.}

\author{Giulio Sciarra}
\email{giulio.sciarra@uniroma1.it}
\affiliation{Dipartimento di Ingegneria Chimica Materiali Ambiente,
             Sapienza Universit\`a di Roma, 
             via Eudossiana 18, I--00184 Roma, Italy}


\begin{abstract}
We study the solutions of a generalized Allen--Cahn equation
deduced from a Landau energy functional, endowed with a non--constant
higher order stiffness. We analytically solve the stationary problem
and deduce the existence of so--called compactons, namely, 
connections on a finite interval between the two phases.
The dynamics problem is numerically solved and compacton formation is
described. 
\end{abstract}




\maketitle

\section{Introduction}
\label{s:introduzione} 
Phase--field models describe physical systems that can exhibit 
different homogeneous phases.  
The state of the system 
on the volume $\Omega\subset\bb{R}^3$ 
is coded into a so called phase--field $u(x,t)$ depending on the 
space and time variables $x\in\Omega$ and $t\in[0,\infty)$, respectively. 
Two values of the phase--field, say $0$ and $1$, represent the two homogeneous 
phases.

These models play a crucial role, for instance, 
in the study of phase reordering
\citep{Bray,Langer,Eyre}:
a system is quenched from the homogeneous high temperature phase into a 
broken--symmetry one (a ferromagnet or a gas abruptly 
cooled below their critical temperature) and the evolution of the 
phase--field $u$ describes the process of separation 
of the two phases.

A straightforward way to derive 
the evolution equation for the field $u$ 
is that of assuming 
a gradient equation \citep{Fife,ABF}
\begin{equation}
\label{grad}
\frac{\partial u}{\partial t} = -\rr{grad}\,H(u)
\end{equation}
associated with 
the Landau energy functional 
\begin{equation}
\label{grad-f}
H(u)
:=
\int_\Omega\Big[\frac{1}{2}\varepsilon\|\nabla u\|^2+W(u)\Big]\rr{d}x
\end{equation}
where $W\in C^2(\bb{R})$ associates an energy with the phase--field $u$ and 
the squared--gradient term of the 
phase--field variations is weighted by the energy cost 
$\varepsilon$, called \textit{higher--order stiffness}. 
According to the usual physical interpretation the energy $W$ has to be 
chosen as a double well function with the two minima 
corresponding to the two phases $0$ and $1$ and $W(0)=W(1)=0$.

If the higher--order stiffness $\varepsilon$ is a constant positive number and 
no constraint to the total value of the field $u$ is imposed, 
it is possible to compute the gradient of the Landau functional 
in the Hilbert space $L^2(\Omega)$
to get the \textit{standard} Allen--Cahn equation 
\begin{equation}
\label{ac}
\frac{\partial u}{\partial t}
=
\varepsilon\Delta u-W'(u)
\end{equation}
with normal derivative of the phase--field on the boundary equal to zero.
Analogously the Allen--Cahn equation endowed with Dirichlet or
mixed boundary conditions could be derived specifying 
a--priori the proper essential
boundary condition in the definition of the Hilbert space in which the gradient of the Landau functional 
should be computed.

The standard Allen--Cahn equation, also called 
the time--dependent Ginzburg--Landau equation, was introduced in \citep{Allen} 
to describe the motion of anti--phase boundaries in crystalline solids.
In this context $u$ represents the concentration of one of the two 
components of the alloy and $\varepsilon$ 
is proportional to the squared interface width. 

In this paper we consider the case in which the higher--order stiffness is 
not constant, but it is a sufficiently regular positive 
function of the field, namely, 
$\varepsilon\in C^2(\bb{R})$ such that $\varepsilon(u)\ge0$ 
for any $u\in\bb{R}$.
This situation has been considered, for instance, in \cite{BSBS}
where the authors studied a similar model to describe glass--like 
relaxation in binary fluid models. 
A closely connected problem, in which a not constant higher--order stiffness 
is used, is the study of the gas--liquid interface in 
capillary tubes \cite{FJ}. 
In this case, the gradient equation \eqref{grad} provides 
the generalized Allen--Cahn equation, see Appendix~\ref{s:ac}, 
\begin{equation}
\label{ac010}
\frac{\partial u}{\partial t}
=
\frac{1}{2}\varepsilon'(u)\|\nabla u\|^2
+\varepsilon(u)\Delta u 
-W'(u)
\end{equation}
again with suitable conditions on the boundary $\partial\Omega$. 

We focus on the one--dimensional case 
$\Omega=[a,b]$
and study the 
solutions of the Allen--Cahn equation \eqref{ac010}
in the case in which the higher--order stiffness coefficient vanishes 
at the phases, namely, $\varepsilon(0)=\varepsilon(1)=0$. 
As we shall discuss in the following section, in such a 
``pathological'' case 
there exist stationary solutions connecting the two phases 
on a finite interval of length $\delta>0$.
It is well known that this 
is not possible in the standard constant higher--order stiffness case, in which 
connections can only be considered on infinite domain \cite{AF}. 

These solutions appeared in the scientific 
literature in different contexts, see, e.g., 
\cite{BSBS,FJ,Witelski}, and have been called \textit{compactons}, in
order to underline the property of being localized within a domain of finite measure. 
Our main goal, here, is to study the behavior of the solutions of 
the evolution equation \eqref{ac010} and, in particular, to describe 
the process leading to the formation of a compacton on the finite 
interval $\Omega=[a,b]$. 
We shall discuss both the interface Dirichlet 
boundary conditions 
\begin{displaymath}
u(a)=0
\;\;\textrm{ and }\;\;
u(b)=1
\end{displaymath}
and the homogeneous Neumann boundary conditions 
\begin{displaymath}
u_x(a)=0
\;\;\textrm{ and }\;\;
u_x(b)=0
\end{displaymath}
We shall respectively refer to these two cases
as to the (D) and (N)--boundary conditions.

Let us summarize our main result. 
Compactons can be used to construct stationary solutions of the 
Allen--Cahn equation performing many excursions between the two 
phases, whose total number is bounded by $(b-a)/\delta$.
In the standard Allen--Cahn stationary problem, i.e., when 
the higher--order stiffness coefficient is constant, stationary profiles
oscillating between the two phases are not allowed when (D)--boundary 
conditions are imposed. On the other
hand, it is possible to construct profiles oscillating between 
two values of the phase--field $u$ ``close'' to the two pure phase 
values in the (N)--boundary condition case. 
These solutions, in the conservative mechanics equivalent 
model language in which the stationary Allen--Cahn model can be 
immediately recasted, correspond to the periodic motions 
of the system with total (kinetic plus potential) energy slightly 
smaller than zero. 

In the (N)--boundary condition case single interface and 
periodic profiles are proven to be unstable \cite{BBB,FH}.
See also \cite{Gurtin}, where it is shown that, in presence 
of a global constraint, the periodic profiles 
are not even local minimizers of the Landau functional \eqref{grad-f} 
defining the model, or 
in other words the corresponding second variation 
of the Landau functional is strictly negative at some perturbation of them. 

We then expect that any time dependent solution of the Allen--Cahn 
evolution equation, for any choice of the initial profile, will 
never tend in the long time limit to one of these oscillating stationary
solutions. In other words, the standard Allen--Cahn evolution 
cannot create an alternating profile and, indeed, such an equation 
is used to model domain coarse--graining in phase separation.

The question we pose in this paper is the following: in presence 
of compactons, can the Allen--Cahn evolution describe the 
alternating profile formation?
In this paper, by means of a numerical computation in the framework of a 
specific model, we shall give a positive answer to such a question.
In particular we shall show that the alternating compacton profile 
formation is possible with both (D) and (N)--boundary conditions. 

In our study we shall use the following techniques: 
the stationary solution of the Allen--Cahn equation \eqref{ac010} 
will be studied analytically and the ``usual'' qualitative 
Weierstrass study will allow the construction of the phase 
portrait which will provide 
a thorough description of the structure of the 
stationary profiles. 
On the other hand, the time--dependent solutions will be studied 
numerically and a code based on the finite element method will 
be adopted.

In order to perform the numerical study a particular 
choice of the functions $\varepsilon(u)$ and $W(u)$ will be done. 
We borrow those functions from 
\cite{FJ} where a model describing the gas--liquid interface in a 
capillary tube has been proposed. 
It is worth noting that we shall not discuss the evolution equations
proposed in \cite{FJ}, but the Allen--Cahn equation with 
stationary profiles coinciding with the ones in the \cite{FJ} model. 
Indeed, our main interest is that of understanding the Allen--Cahn 
evolution in presence of compactons and to this aim we have 
chosen, as a prototype model, the one in \cite{FJ} whose stationary 
solutions has a clear physical interpretation. 
Moreover, this model allows to study analytically the compactons, 
whose behavior can be expressed in terms of special functions. This 
will provide us with an effective analytical control of our 
numerical results. 

One of the main 
results in \cite{FJ} is the possibility to describe the 
existence of local, non--spreading, and compactly supported 
bubbles in a capillary tube. 
In that paper the model was studied numerically. Here we solve analytically
the equation giving the stationary states of the system 
and explain some of the 
features of the compacton solutions presented in \cite{FJ}. 

The paper is organized as follows: in Section~\ref{s:compattoni}
we discuss under quite general hypotheses the existence 
of compactons for the stationary Allen--Cahn equations.
In Section~\ref{s:modello} we consider the model introduced in \cite{FJ}
to study the gas--liquid interface in capillary tube and, in such a 
context, we find explicitly (in terms of special function) 
the compacton solutions and discuss their main physical properties.
In Section~\ref{s:dinamica} we study numerically the solutions 
of the Allen--Cahn equation with higher--order stiffness $\varepsilon$ 
and energy $W$ as in \cite{FJ}.
Section~\ref{s:conclusioni} is devoted to some brief conclusions. 

\section{Compactons}
\label{s:compattoni}
We consider the Allen--Cahn problem \eqref{ac010} 
on the one--dimensional space $[a,b]$. The equation 
for the stationary solutions $u=u(x)$
then becomes
\begin{equation}
\label{ac020}
\varepsilon(u)u_{xx}
+\frac{1}{2}\varepsilon'(u)u_x^2
-W'(u)
=0
\;\;.
\end{equation}
Here, and in the following, the prime will always denote the 
derivative with respect to the natural argument, whereas space (time)
derivatives will be written explicitly as $\rr{d}/\rr{d} x$  or 
with a subscript $x$ ($\rr{d}/\rr{d} t$ or with a subscript $t$). 

We assume $W(u)=W_0u^2(1-u)^2$, with $W_0>0$, and 
$\varepsilon\in C^2([0,1])$ such that 
$\varepsilon(u)>0$ for $u\in(0,1)$, and 
$\varepsilon(0)=\varepsilon(1)=0$.
The choice of the potential energy $W$ with two isoenergetic minima
models the existence of two coexisting phases. 
Moreover, we assume that 
$\varepsilon$ tends to $0$ in $0$ and $1$ at least as a power law, namely,
there exists $\chi>0$ such that 
\begin{displaymath}
\lim_{u\to0^+}\frac{\varepsilon(u)}{u^\chi}=0 
\,\textrm{ and } 
\lim_{u\to1^-}\frac{\varepsilon(u)}{(1-u)^\chi}=0
\;\;.
\end{displaymath}
The two last assumptions are crucial for the compacton existence\footnote{We
have chosen the Duffing potential energy $W$ for simplicity. The same 
discussion can be repeated for very general double well potential 
energies, but the condition on the higher--order 
stiffness coefficient have to be 
chosen accordingly.}, see \eqref{inter20} and the discussion which follows, as well as the related
arguments in \cite{BSBS,FJ}.

It is very important to remark that any regular solution 
$u(x)$ of \eqref{inter00} is such that the conservation law 
\begin{equation}
\label{inter10}
\frac{\rr{d}}{\rr{d}x}
\Big[
     \frac{1}{2}\varepsilon(u)\,u_x^2
     -W(u)
\Big]
=0
\end{equation} 
is satisfied. 

Note that the problem is similar, see also 
\cite{CI,CIS2010,CIS2012,CIS2013}, to that of an holonomic conservative 
mechanical system with Lagrangian coordinate $u$, not constant 
mass matrix $\varepsilon(u)$, and potential energy of the conservative 
force $-W(u)$, once the space variable $x$ is interpreted as time.
A lot of care has to be used when one wants to exploit 
this analogy, since the mass matrix $\varepsilon(u)$ is not positive 
defined, but it is equal to zero in the pure phases 
$u=0$ and $u=1$. 

The aim of this model is that of describing a compact 
interface (or connection) 
between the pure phases $u=0$ and $u=1$, namely, 
we look for a solution of \eqref{ac020} equal to zero on a finite 
(say left) space interval, equal to one on a finite (say right) 
space interval, and continuously joining these two pure phases 
on an intermediate ``finite" space interval. 
This intermediate interval will be the compact interface 
(or connection)
between the two pure phases. 

In standard cases, i.e., when the higher--order 
stiffness coefficient is constant, 
an interface with zero 
derivative at the boundary can only be achieved on an infinite 
space interval (heteroclinic problem). This property is very general 
and is connected to the uniqueness of the solution of a Cauchy problem 
which is ensured if the differential equation describing the interface 
is sufficiently regular. 
In the model we are studying here, this regularity of the equation is 
not satisfied due to the presence of the not positive definite 
mass matrix $\varepsilon(u)$. This is the key peculiarity of the model 
that gives rise to the existence of compacton solutions. 

First of all we note that the constant functions $u(x)=0$
and $u(x)=1$ trivially satisfy \eqref{ac020}. So that 
we can imagine to construct a solution of this equation 
such that $u(x)=0$ for all $x\in[a,c]$ and 
$u(x)=1$ for all $x\in[c+\delta,b]$, with $c,\delta\in\bb{R}$ given. 
The problem, now, is that of finding the interface joining the 
two pure phases on the ``finite" interval $[c,c+\delta]$. 
Note that the pure phases fix the value of the constant of motion 
\eqref{inter10} to zero; hence, the interface we are looking for has to 
satisfy 
\begin{displaymath}
\frac{1}{2}\varepsilon(u(x))\,u_x^2
-W(u(x))=0
\;\;\textrm{ for } x\in[c,c+\delta]
\end{displaymath}

By separation of variables we get the implicit solution
\begin{equation}
\label{inter20}
x-c
=
\int_0^{u}\!\!\rr{d}y\,\sqrt{\frac{\varepsilon(y)}{2 W(y)}}
\end{equation}
Since we assumed $\varepsilon$ to vanish in $0$ and $1$ at least 
as a power, we have that the integral above is convergent 
on the interval $[0,1]$. Hence, we have proven the existence 
of the compacton and we can also conclude that 
\begin{equation}
\label{inter25}
\delta
=
\int_0^1\!\!\rr{d}u\,\sqrt{\frac{\varepsilon(u)}{2 W(u)}}
\end{equation}
expresses its width. 

We close this section by noting that, by means of the 
conservation law, it is possible to describe the structure of all the 
solutions of the stationary equation \eqref{ac020}. Indeed, 
\eqref{inter10} ensures that any regular solution satisfies the 
equation
\begin{equation}
\label{inter100}
     \frac{1}{2}\varepsilon(u)\,u_x^2
     -W(u)
=E
\end{equation}
for some $E\in\bb{R}$.

The structure of the solution of the equation \eqref{inter100} 
lying in the interval $0\le u\le 1$ is as follows. 
For $E=0$ the constant, $u(x)=0$ and $u(x)=1$, solutions and
combination of them with compactons are found. Note that, since we assumed 
a power law behavior of $\varepsilon(u)$ for $u\to0,1$, we have that 
the space derivative of the profile can be zero, finite, or divergent 
in the phases $0$ and $1$. For $E>0$, since 
$\varepsilon(u)$ vanishes in $u=0$ and $u=1$, the profile $u(x)$ must have 
divergent derivative in $0$ and $1$. 
For $-W_0/8<E<0$ the solution 
is bounded to the region in which $W(u)+E\ge0$; since in such a region 
$\varepsilon(u)>0$, we find a classical oscillating solution. 
Finally, for $E=-W_0/8$, the unique solution 
is the constant $u(x)=1/2$.

\begin{figure}[t]
\begin{picture}(200,120)(-60,-10)
\thinlines
\put(0,0){\vector(0,1){100}}
\put(-5,50){\vector(1,0){120}}
\put(0,50){\circle*{5}}
\put(50,50){\circle*{5}}
\put(100,50){\circle*{5}}
\thicklines
\qbezier[30](0,60)(50,75)(100,60)
\qbezier[30](0,40)(50,25)(100,40)
\thinlines
\qbezier(0,100)(0,75)(50,75)
\qbezier(51,75)(100,75)(100,100)
\qbezier(0,0)(0,25)(50,25)
\qbezier(51,25)(100,25)(100,0)

\qbezier(20,50)(20,60)(50,60)
\qbezier(50,60)(80,60)(80,50)
\qbezier(20,50)(20,40)(50,40)
\qbezier(50,40)(80,40)(80,50)

\put(115,43){${u}$}
\put(-12,100){${u_x}$}
\end{picture}
\caption{A possible phase portrait of the 
stationary equation \eqref{ac020}.
The dotted lines represent the compactons: in the picture 
they assume finite values at $u=0$ and $u=1$, but, 
recall, they could also tend to zero or diverge
(see, also, figure \ref{f:qual-ju}). 
}
\label{f:qual}
\end{figure}
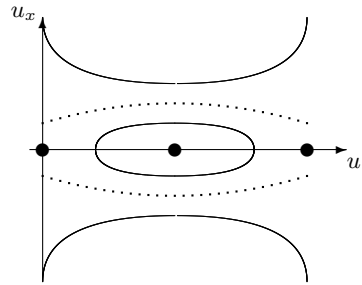

These results are summarized in the figure~\ref{f:qual} in which 
the three points represent the constant solutions $u(x)=0$, $u(x)=1/2$,
and $u(x)=1$, 
the dotted lines represent the compactons, 
the lines diverging in $0$ and $1$ are the solutions for $E>0$, and, 
finally, the closed loops are the solutions for $-W_0/8<E<0$.
Note that in the figure we have depicted the compacton line 
finite at the phases, but, as we discussed above, it can happen 
that close to the phases the line tends to zero or diverges. 

Recalling that $\delta$ denotes the length of the compactons solution, 
note that the length 
\begin{displaymath} 
\delta_\textrm{u}(E)
=
\int_0^{1}\!\!\rr{d}y\,\sqrt{\frac{\varepsilon(y)}{2[E+W(y)]}}
\end{displaymath} 
of profiles connecting the two phases $0$ and $1$ 
and corresponding to $E>0$ 
is such that 
\begin{displaymath}
\delta_\textrm{u}(E)<\delta
\;\;\textrm{ for }\;\;E>0
\;\;.
\end{displaymath}
On the other hand, for $E<0$ we let $0<u_-(E)<u_+(E)<1$ be the two 
solutions of the equation $W(u)+E=0$ lying in the open interval 
$(0,1)$. Hence, the length of a single interface connecting $u_-(E)$ to 
$u_+(E)$ is given by
\begin{displaymath} 
\delta_\textrm{d}(E)
=
\int_{u_-(E)}^{u_+(E)}\!\!\rr{d}y\,\sqrt{\frac{\varepsilon(y)}{2[E+W(y)]}}
\end{displaymath} 

This analysis on the phase space trajectories
allows us to state the 
following results about the existence of solutions of the stationary problem. 
The stationary equation \eqref{ac020} with (D)--boundary conditions
has a unique solution corresponding to a phase line with $E>0$ 
if $b-a<\delta$, 
has the unique compacton solution if $b-a=\delta$, 
and has infinite solutions if $b-a>\delta$ which can be constructed 
by gluing compactons and pure phase constant segments.

The stationary equation \eqref{ac020} with (N)--boundary conditions
has always the two pure phase constant solutions $u(x)=0$ and $u(x)=1$. 
If $b-a$ is large enough, so that for some $E<0$ one has 
$\delta_\textrm{d}(E)<b-a$, 
the problem can have single connection or 
oscillating solutions connecting two points $0<u_-<u_+<1$ and 
corresponding to the phase lines with $E<0$. 
Moreover, if 
$b-a>\delta$ the problem has also 
solutions which can be constructed 
by gluing compactons and pure phase constant segments.

\section{Bubbles in a capillary tube}
\label{s:modello}
A one dimensional model is adopted for describing the spatial distribution, in a capillary tube, of the liquid and the gaseous phases
regarding the mixture as a non--uniform fluid, which means, 
according to \cite{CH_I}, 
a system having a spatial variation of one of its intensive scalar 
properties. In particular following \cite{FJ} one can assume this property, say the phase--field introduced in section \ref{s:compattoni},
to be the density of the gas with respect to the volume locally available.
In the specific case of a capillary tube with a constant section, the 
phase--field 
is the fraction of the cross--sectional area of the tube 
occupied by the gaseous phase $S_{\rr g}$, per unit length of the tube.

Apparently the gas saturation $S_{\rr g}$ can be related to the volume 
density of the liquid phase $S_{\rr l}$ keeping in mind the obvious constraint 
\begin{equation}
\label{vincolo}
S_\rr{g}+S_\rr{l}=1.
\end{equation}

According with the general formulation presented in section \ref{s:compattoni}, a Landau energy functional 
is introduced whose density per unit volume is the sum of a bulk contribution, prescribed in terms of a double well potential,
$F(S_{\rr g})$, and an energy penalty
for gradients of the gas saturation $S_{\rr g}$, 
affected by the current value of $S_{\rr g}$.
In order to characterize the admissible equilibrium configurations of the system we refer from now on to the 
constitutive model given in \cite{FJ}. Assuming the equilibrium between the gaseous and the liquid phase to be controlled only by 
capillary forces and therefore by the adjustment of the 
contact angle $\theta\in(0,\pi)$ between the gas--liquid and the 
liquid--solid interfaces, see \cite{deGennes}, the double well potential $F(S_{\rr g})$ is prescribed following
\cite{FJ} by
\begin{equation}
\label{total_bulk_energy}
\begin{array}{rcl}
F(S_{\rr g})
&\!\!=&\!\!
\dfrac{\gamma\, \left(1-\, \cos \theta \right)}{R} \left(1-S_{\rr g} \right)^2 S_{\rr g}^2\medskip
\\
&&\!\!
+\dfrac{\gamma\, \cos \theta}{R} \left[\left(1-S_{\rr g} \right)^2 -S_{\rr g}^2 \right],
\\
\end{array}
\end{equation}
$\gamma$ being the surface energy relative to the gas--liquid interface, and $R$ the radius of the capillary tube. 
Following \cite{FJ} the higher--order stiffness will be written in terms of 
\begin{equation}
\label{free05}
\begin{array}{l}
\Gamma=C_{\Gamma}\,\gamma R\, (1-\cos\theta)
       \Big[\dfrac{1-\sin\theta}{\cos\theta}\Big]^2
\end{array}
\end{equation}
and
\begin{equation}
\label{free06}
\begin{array}{l}
k(S_\rr{g})=S_\rr{g}^\alpha (1-S_\rr{g})^\beta,
\end{array}
\end{equation}
see \eqref{staz}, with
$\alpha=2-(1/2)\cos\theta$ and 
$\beta=2+(1/2)\cos\theta$.
In \cite{FJ} a dimension argument is given for the definition of $\Gamma$,
moreover, 
it is remarked that the peculiar expression of $k$ plays 
a key role in the existence of compact interfaces, see also 
\cite{BSBS}.

The derivative of the double well potential \eqref{total_bulk_energy} specifies
the difference between the chemical potential of the gas and the chemical potential of the liquid or, 
analogously, the negative chemical potential $\mu$ of the liquid, once that of the gas has been fixed to zero, as a reference value. 
Its value at the pure phases, $S_{\rr g}=1$, the gas, and $S_{\rr g}=0$, the liquid, is the same, say
\begin{equation}
\label{chem_pot}
\left.\dfrac{\partial F}{\partial S_{\rr g}}\right|_{ S_{\rr g}=0}=\left.\dfrac{\partial F}{\partial S_{\rr g}}\right|_{ S_{\rr g}=1}=-2 \dfrac{\gamma \cos \theta}{R},
\end{equation}
so that, according with classical Maxwell's rule, the 
non--uniform fluid exhibits coexistence of the two phases at equilibrium only when
the chemical potential is uniformly equal to
$\mu=\partial F/\partial S_{\rr g}=-2\, \gamma \cos \theta/R$ over the whole spatial domain.
Requiring this condition to be verified 
corresponds to find out the solutions of the minimization
problem
\begin{equation}
\label{minimum_bulk_energy}
\underset{S_{\rr g}}{\rr {min}}\left( F(S_{\rr g})+2\, \dfrac{\gamma \cos \theta}{R}\, S_{\rr g}\right),
\end{equation}
which admits two solutions at $S_{\rr g}=0$ and  $S_{\rr g}=1$. Due to the additional linear term, $2\, \gamma \cos \theta /R\, S_{\rr g}$ the two phases 
correspond, now, to two
isopotential minima of the 
function $\mathcal F(S_{\rr g})=F(S_{\rr g})+2\, \gamma \cos \theta /R\, S_{\rr g}$. 

The regularization provided by the energy penalty proportional to the squared--gradient term via the higher--order stiffness $\Gamma k(S_{\rr g})$, see 
equations \eqref{free05}--\eqref{free06}, implies,
at coexistence, the conservation law \eqref{inter10} to be rewritten as follows:
\begin{equation}
\label{staz}
\begin{array}{rl}
 0 =& \dfrac{2\gamma}{R}(1-\cos\theta)(1-S_{\rr g})S_{\rr g}(1-2S_{\rr g})
+\medskip\\
  &{\displaystyle
 -\Gamma 
\sqrt{k(S_{\rr g})}
\,
\frac{\rr{d}}{\rr{d}x}
\left(\sqrt{k(S_{\rr g})}
\,
\frac{\rr{d}}{\rr{d}x}
S_{\rr g} \right)
},
\end{array}
\end{equation}
which therefore reads as a specialization of the Allen--Cahn equation when a non--uniform fluid is placed into a capillary tube.

\subsection{The compact interface problem}
\label{s:interfaccia}
From now on we shall simplify the notation by letting 
$S_\rr{g}=S$ and rewrite 
equation \eqref{staz} as 
\begin{equation}
\label{inter00}
\Gamma k(S)
\,S_{xx}
+
\frac{1}{2}\Gamma
S_x^2
k'(S)
-
V'(S)
=0
\end{equation}
where we have set 
\begin{equation}
\label{inter06}
\begin{array}{rcl}
V(S)
&\!\!=&\!\!
{\displaystyle
 F(S)+2S\frac{\gamma}{R}\cos\theta-\frac{\gamma}{R}\cos\theta
 \vphantom{\bigg\{_\}}
}
\\
&\!\!=&\!\!
{\displaystyle
\frac{\gamma}{R}(1-\cos\theta)(1-S)^2S^2
}
\end{array}
\end{equation}
In order for the physical dimensions of the quantities introduced above to be consistent with the
notation of Section~\ref{s:introduzione}
a suitable viscosity parameter $\mu$ must be introduced so that $\mu W=F$ and $\varepsilon=\Gamma k/\mu$.

It is important to remark that the interface problem \eqref{inter00}
in the 
capillarity setup is an example of applications of 
the theory developed in Section~\ref{s:compattoni}.
Thus, 
as discussed in Section~\ref{s:compattoni}, 
in order to ensure that the integral \eqref{inter20} 
is convergent, it is sufficient to require that the parameters 
$\alpha,\beta$ in \eqref{free06} are strictly positive. 
In other words the particular dependence of $\alpha$ and $\beta$ 
on the contact angle $\theta$ 
discussed below \eqref{free06} is not necessary to prove 
the existence of compactons, but, as we shall see below, 
affects their width $\delta$.
Indeed, by \eqref{inter25}
we get 
\begin{equation}
\label{inter27}
\delta
=
\int_0^1\frac{y^{-(1/4)\cos\theta}(1-y)^{(1/4)\cos\theta}}
                 {\sqrt{2\gamma(1-\cos\theta)/(R\Gamma)}}
\,\,\rr{d}y
\end{equation}
for the compacton width.
Moreover, by \eqref{app60} we have 
\begin{displaymath}
\delta
=
\frac{1}{\sqrt{2\gamma(1-\cos\theta)/(R\Gamma)}}
\,
\frac{(\pi/4)\cos\theta}{\sin[(\pi/4)\cos\theta]}
\end{displaymath}
Finally, recalling \eqref{free05}, we get 
\begin{equation}
\label{inter30}
\delta
=
R\sqrt{\frac{C_\Gamma}{2}}
\,\frac{1-\sin\theta}{|\cos\theta|}
\,\frac{(\pi/4)\cos\theta}{\sin[(\pi/4)\cos\theta]}
\end{equation}

\subsection{Compacton profile}
\label{s:profile}
As above it is possible to write an implicit expression of 
the compacton profile $S(x)$ in terms of special functions. 
Indeed, by performing the same computation as above, 
from \eqref{inter20} we get 
\begin{equation}
\label{inter32}
x(S)-c
=
\int_0^S\frac{y^{-(1/4)\cos\theta}(1-y)^{(1/4)\cos\theta}}
                 {\sqrt{2\gamma(1-\cos\theta)/(R\Gamma)}}
\,\,\rr{d}y
\end{equation}
Equation \eqref{app50}
and some simple algebra yields 
\begin{equation}
\label{inter35}
\begin{array}{rcl}
x(S)-c
&\!\!=&\!\!
{\displaystyle
 R\sqrt{\frac{C_\Gamma}{2}}
 \,\frac{1-\sin\theta}{|\cos\theta|}
 \vphantom{\bigg\{_\}}
}\\
&&\!\!
{\displaystyle
 \times
 B\Big(S,1-\frac{1}{4}\cos\theta,1+\frac{1}{4}\cos\theta\Big)
}\\
\end{array}
\end{equation}
where we have denoted by $B$ the incomplete beta function, see
\eqref{app10} in Appendix~\ref{s:integrali},
which 
gives implicitly the profile of the compacton $S(x)$ for 
$x\in(c,c+\delta)$. 

By using the explicit solution given above, many interesting 
physics feature of the compactons discussed in \cite{FJ} can be 
proven analytically. For instance, 
in that paper it has been noted that the convexity of the 
interface profile $S(x)$ for $x\in[c,c+\delta]$ depends on 
whether the liquid phase has a 
wetting ($\theta>\pi/2$, for instance water) 
or 
a not wetting ($\theta<\pi/2$, for instance mercury) 
behavior.
By means of \eqref{inter35} 
this problem is reduced to a simple computation. Indeed, 
recall that
the compacton satisfies \eqref{inter00} and along 
the compacton the constant of motion \eqref{inter10} is equal 
to zero; thus, from \eqref{inter00} and \eqref{inter10}, we get 
that 
\begin{displaymath}
\Gamma k(S)\,S_{xx}
=V(S)\Big[\frac{1}{V(S)}V'(S)
          -
          \frac{1}{k(S)}k'(S)\Big]
\end{displaymath}
for any $x\in(c,c+\delta)$. 
A simple computation yields
\begin{displaymath}
\frac{1}{V(S)}V'(S)
=
\frac{2(1-2S)}{S(1-S)}
\end{displaymath}
and 
\begin{displaymath}
\frac{1}{k(S)}k'(S)
=
\frac{\alpha-4S}{S(1-S)}
\end{displaymath}
Thus, for any $x\in(c,c+\delta)$ we have that 
\begin{equation}
\label{inter40}
S_{xx}(x)
=
\frac{V(S(x))}{2\Gamma S(x)[1-S(x)]k(S(x))}
\,\cos\theta
\end{equation}
Since, $\Gamma\ge0$ and 
$V(S(x))$, $S(x)$, $1-S(x)$, and $k(S(x))$ are strictly positive
in the open interval $(c,c+\delta)$,  
we have that the profile is convex for $\theta<\pi/2$ (not wetting 
liquid) and not convex for $\theta>\pi/2$ (wetting liquid).  

\begin{figure}
\begin{picture}(100,120)
\put(-70,-30)
{
\resizebox{8cm}{!}{\rotatebox{0}{\includegraphics{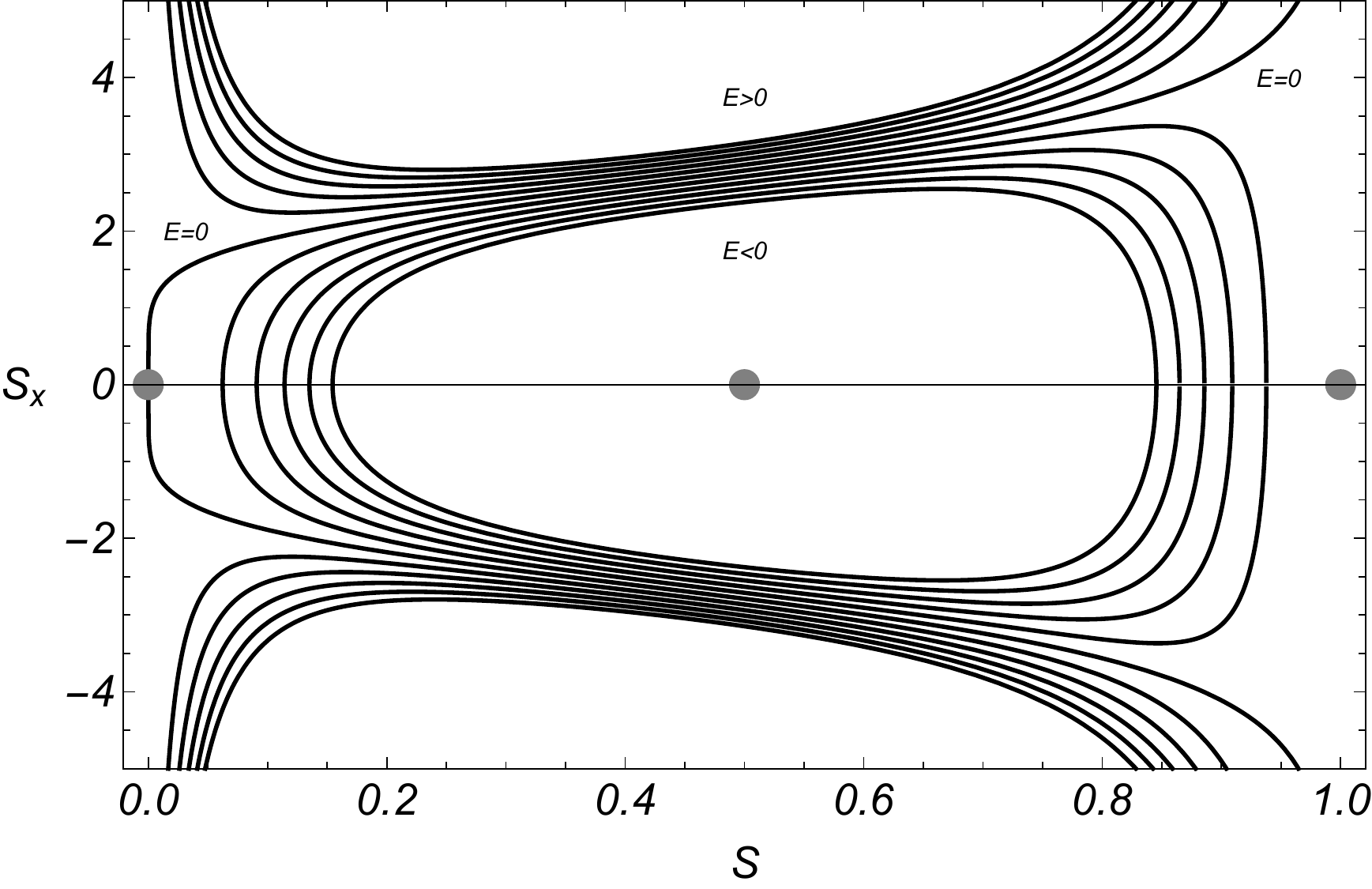}}}
}
\end{picture}  
\vskip 1. cm
\caption{
Phase portrait associated with the equation 
\eqref{port000} 
for $\gamma=1$, $R=1$, $C_\Gamma=3/2$, and $\theta=\pi/4$.}
\label{f:qual-ju}
\end{figure}

\subsection{Phase portrait}
\label{s:ritratto}
In this section we discuss the structure of the solutions of 
the stationary equation \eqref{inter00} by means of the 
qualitative Weierstrass analysis. 
The conservation law \eqref{inter10} in this case reads
\begin{equation}
\label{port000}
\frac{1}{2}\Gamma k(S)
S_x^2
-V(S)
=E
\end{equation}
with $E\in\bb{R}$.

The phase portrait of the model can be deduced by solving 
\eqref{port000} with respect to $S_x$. 
For $\gamma=1$, $R=1$, $C_\Gamma=3/2$, and $\theta=\pi/4$ we find the 
drawing depicted in figure~\ref{f:qual-ju}.
The disks in the pictures denotes the constant solution, the line 
tending to zero in zero represents the compacton, closed 
curves are associated to the cases $E<0$, the remaining lines 
represent the profiles in the case $E>0$.

\begin{figure}
\begin{picture}(100,120)
\put(-70,-30)
{
\resizebox{8cm}{!}{\rotatebox{0}{\includegraphics{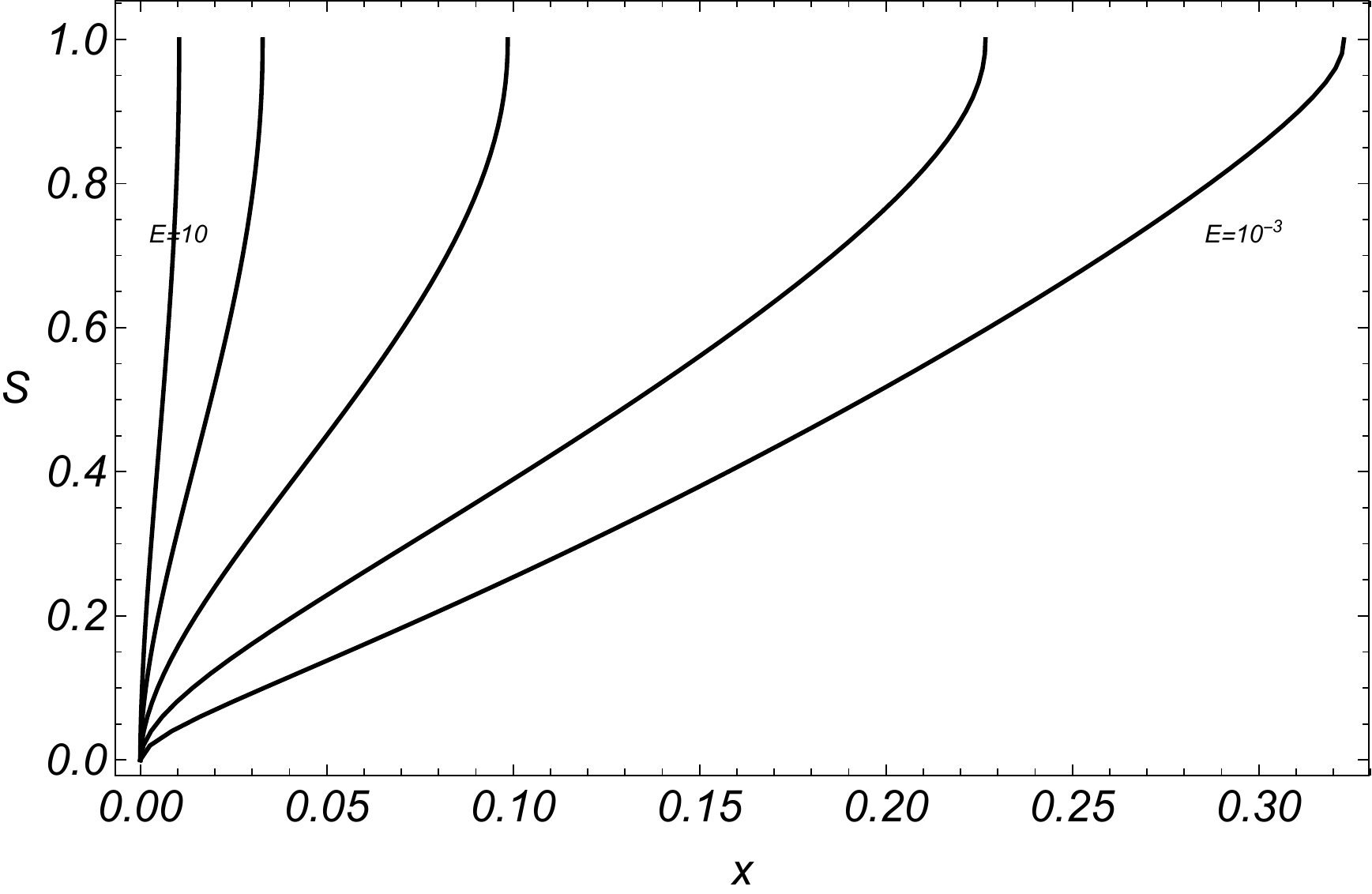}}}
}
\end{picture}  
\vskip 1. cm
\caption{From the left to the right, 
we plot the stationary profiles of the equation \eqref{inter00} 
computed via the integral \eqref{port010} for $E=10,1,0.1,0.01,0.001$.
Parameters: 
$\gamma=1$, $R=1$, $C_\Gamma=3/2$, and $\theta=\pi/4$.}
\label{f:prof}
\end{figure}

\begin{figure}
\begin{picture}(100,120)
\put(-70,-30)
{
\resizebox{8cm}{!}{\rotatebox{0}{\includegraphics{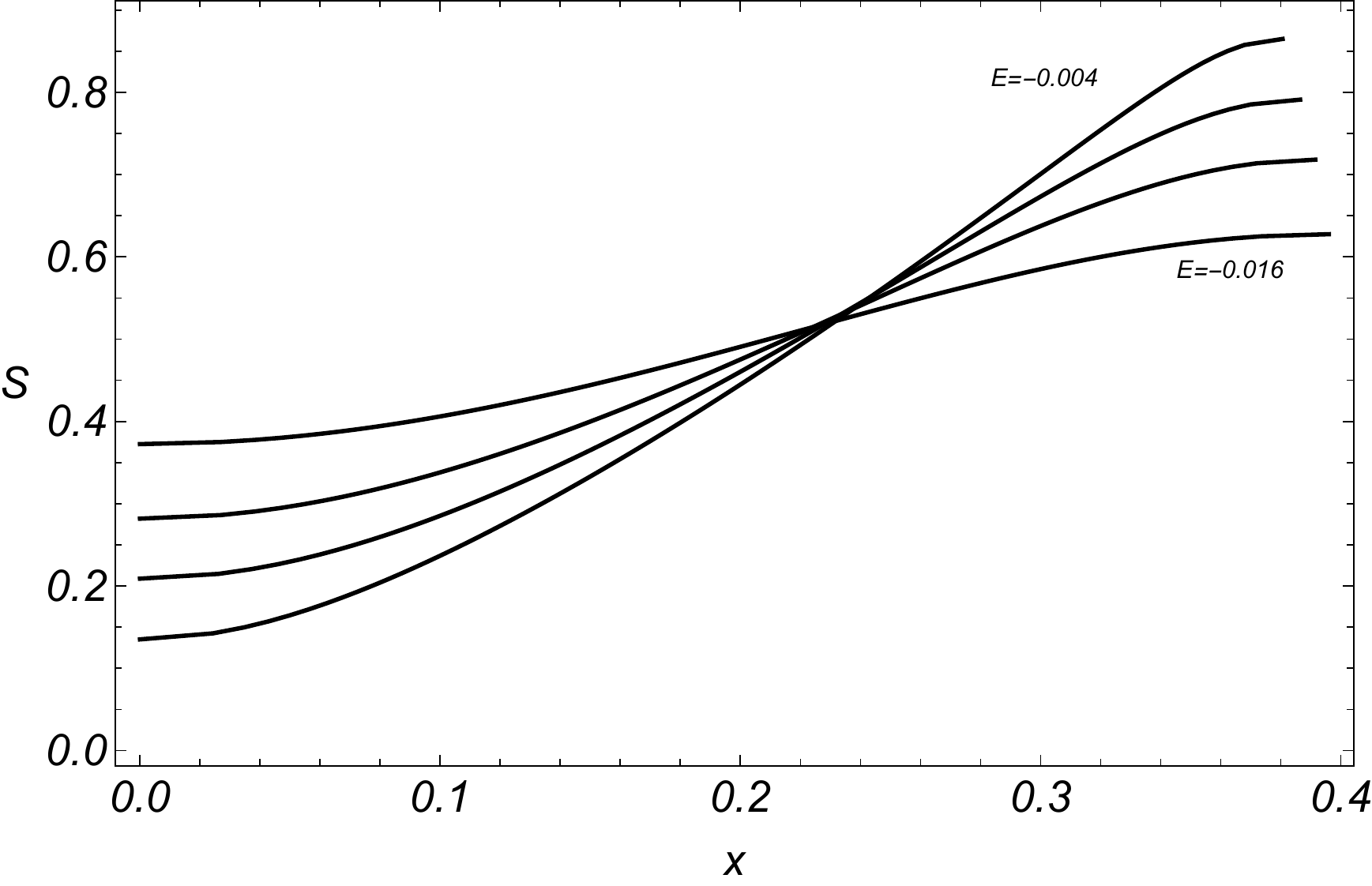}}}
}
\end{picture}  
\vskip 1. cm
\caption{We plot the stationary profiles of the equation \eqref{inter00} 
computed via the integral \eqref{port020} for 
$E=-0.016,-0.012,-0.008,-0.004$.
Parameters: 
$\gamma=1$, $R=1$, $C_\Gamma=3/2$, and $\theta=\pi/4$.}
\label{f:prof-N}
\end{figure}

In order to find the stationary profiles one has to integrate 
the equation \eqref{port000}.
For $E=0$ the solutions of \eqref{port000} are the 
constant profiles $S(x)=0$ and $S(x)=1$, and the 
compacton. 
For $E>0$, the problem of finding the stationary 
profiles (in an implicit form) is reduced to the computation of the  
definite integral
\begin{equation}
\label{port010}
x=\int_{0}^{S(x)}\sqrt{\frac{\Gamma k(y)}{2[E+V(y)]}}\,\rr{d}y
\end{equation}
see the figure~\ref{f:prof}.
For $E<0$, denoted by $S_-<S_+$ the two solutions 
of the equation $V(S)+E=0$ lying in the open interval $(0,1)$,
the problem of finding the stationary 
profiles (in an implicit form) is reduced to the computation of the  
definite integral
\begin{equation}
\label{port020}
x=\int_{S_-}^{S(x)}\sqrt{\frac{\Gamma k(y)}{2[E+V(y)]}}\,\rr{d}y
\end{equation}
see the figure~\ref{f:prof-N}.

\section{Approaching compactons}
\label{s:dinamica}
Once defined the admissible stationary configurations, which solve \eqref{inter100},
in particular
those describing the spatial distribution of the liquid and the gaseous phases in a
capillary tube, see \eqref{inter35}, \eqref{port010}, and \eqref{port020} and figures~\ref{f:prof} and \ref{f:prof-N}, it is interesting to discuss
which of them can be attained through the dissipative evolution described by the Allen--Cahn
equation \eqref{ac010}, endowed with (D) or (N)--boundary conditions. 

In the following the solutions of the Allen--Cahn equation with (D) and (N)--boundary conditions are separately discussed when considering
$(b-a)<\delta$, say the length of the interval smaller than the length of the compacton, and $(b-a)>\delta$.
In the first case no compacton stationary profile is admissible, conversely in the second one
suitable profiles, constructed gluing compactons and pure phases, are 
admissible solutions of the problem. The time--dependent spatial profiles
are numerically captured using a finite element code which has been implemented within 
MATHEMATICA. Time is made dimensionless with respect to
the ratio $\mu/(\gamma/R)$.

\begin{figure}
\begin{picture}(100,120)
\put(-70,-30)
{
\resizebox{8cm}{!}{\rotatebox{0}{\includegraphics{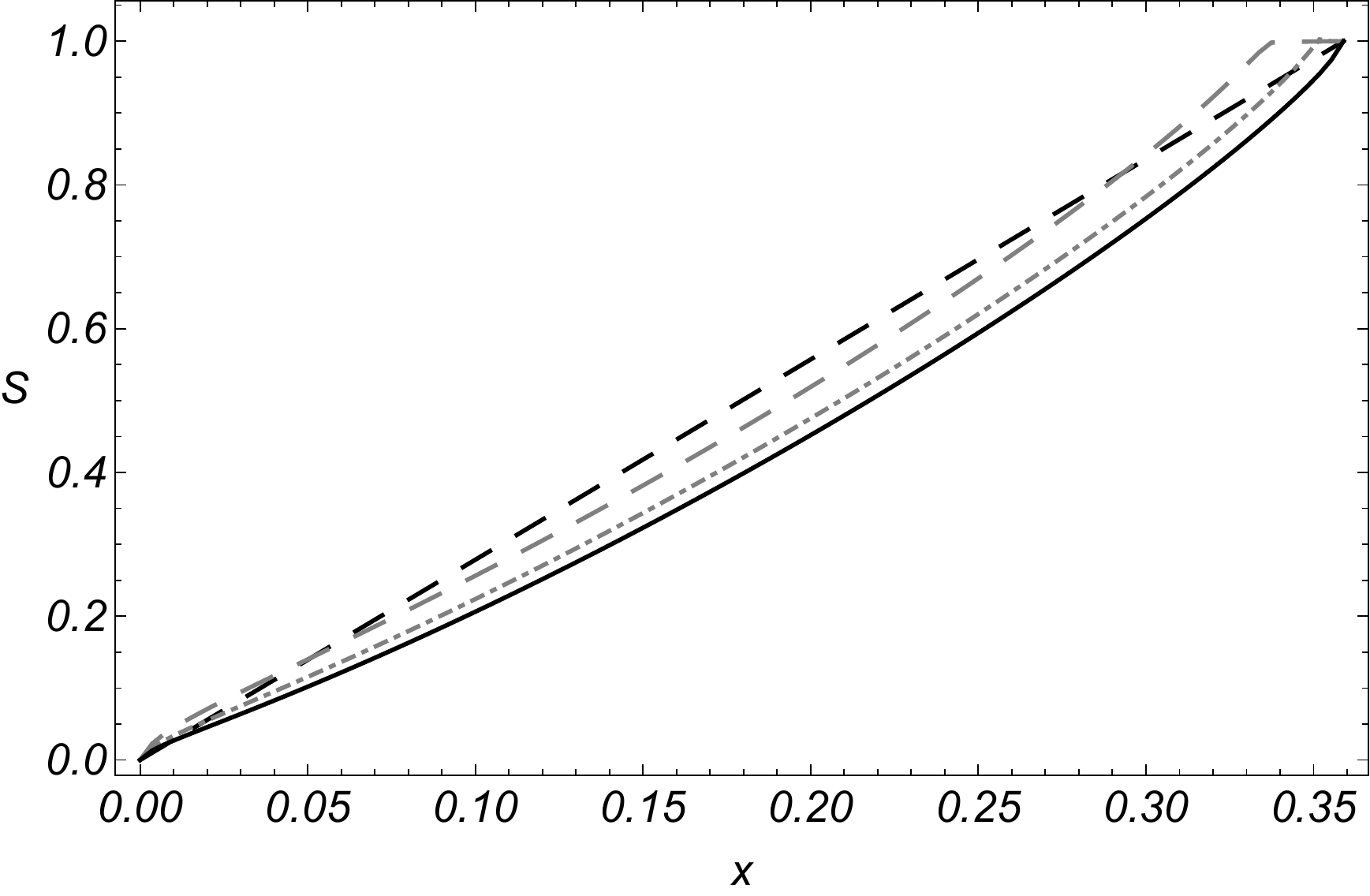}}}
}
\end{picture}  
\vskip 1. cm
\caption{Allen--Cahn dynamics with (D)--boundary conditions
and a linear initial profile (dashed black) connecting
the two phases. Two intermediate profiles (dashed gray) and the 
stationary profile (solid black) are depicted.
The domain of length $(b-a)=19/20\, \delta$ is discretized into $10^2$ finite elements, the 
time needed 
to get a distance of $10^{-3}$ between two subsequent profiles is $t_f=189$.}
\label{f:Dbcs_L<delta}
\end{figure}

\begin{figure}
\begin{picture}(100,120)
\put(-70,-30)
{
\resizebox{8cm}{!}{\rotatebox{0}{\includegraphics{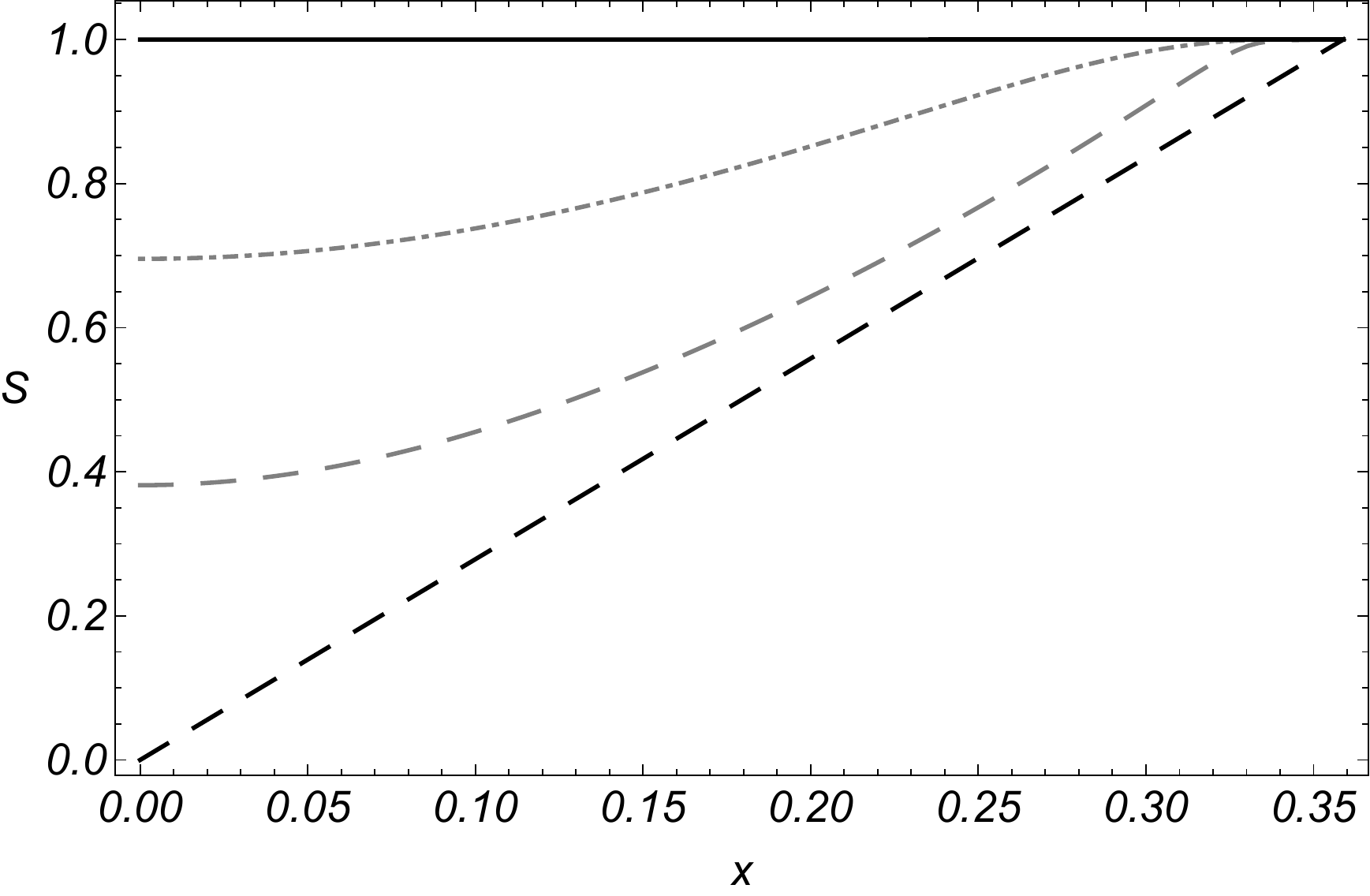}}}
}
\end{picture}  
\vskip 1. cm
\caption{Allen--Cahn dynamics with (N)--boundary conditions
and a linear initial profile (dashed black) connecting
the two phases. Two intermediate profiles (dashed gray) and the 
stationary profile (solid black) are depicted. 
The domain of length $(b-a)=19/20\, \delta$ is discretized into $10^2$ finite elements, the 
time needed 
to get a distance of $10^{-3}$ between two subsequent profiles is $t_f=32$.}
\label{f:Nbcs_L<delta}
\end{figure}

Let $(b-a)<\delta$, in this case the dynamics does not tend to the compacton simply because there 
is not sufficient space for the compacton to arise. Assuming (D)--boundary conditions,
the stationary configuration is a regular profile, see figure~\ref{f:Dbcs_L<delta}, 
whilst for (N)--boundary conditions the dynamics tends to one of the two
pure phases, depending on the initial data, see figure~\ref{f:Nbcs_L<delta}. 
It is interesting to notice that intermediate profiles of $S$,
for (D)--boundary conditions, can be obtained gluing regular profiles, of length smaller than 
$(b-a)$, similar to those of figure~\ref{f:prof}, and pure phase solutions (in particular $S=1$),
where the measure of the subdomain corresponding to this last partial solution fades away with 
increasing time. On the other hand assuming (N)--boundary conditions the evolution passes through
a progressive flattening of the profiles.

Consider now $(b-a)>\delta$, for both (D) and (N)--boundary conditions
two different situations are discussed corresponding to a length of the interval 
$(b-a)$ larger or much larger than $\delta$. In the first case only one compacton can 
arise, whilst in the second one more than one compacton can form, depending on the initial conditions.
\begin{figure}
\begin{picture}(100,120)
\put(-70,-30)
{
\resizebox{8cm}{!}{\rotatebox{0}{\includegraphics{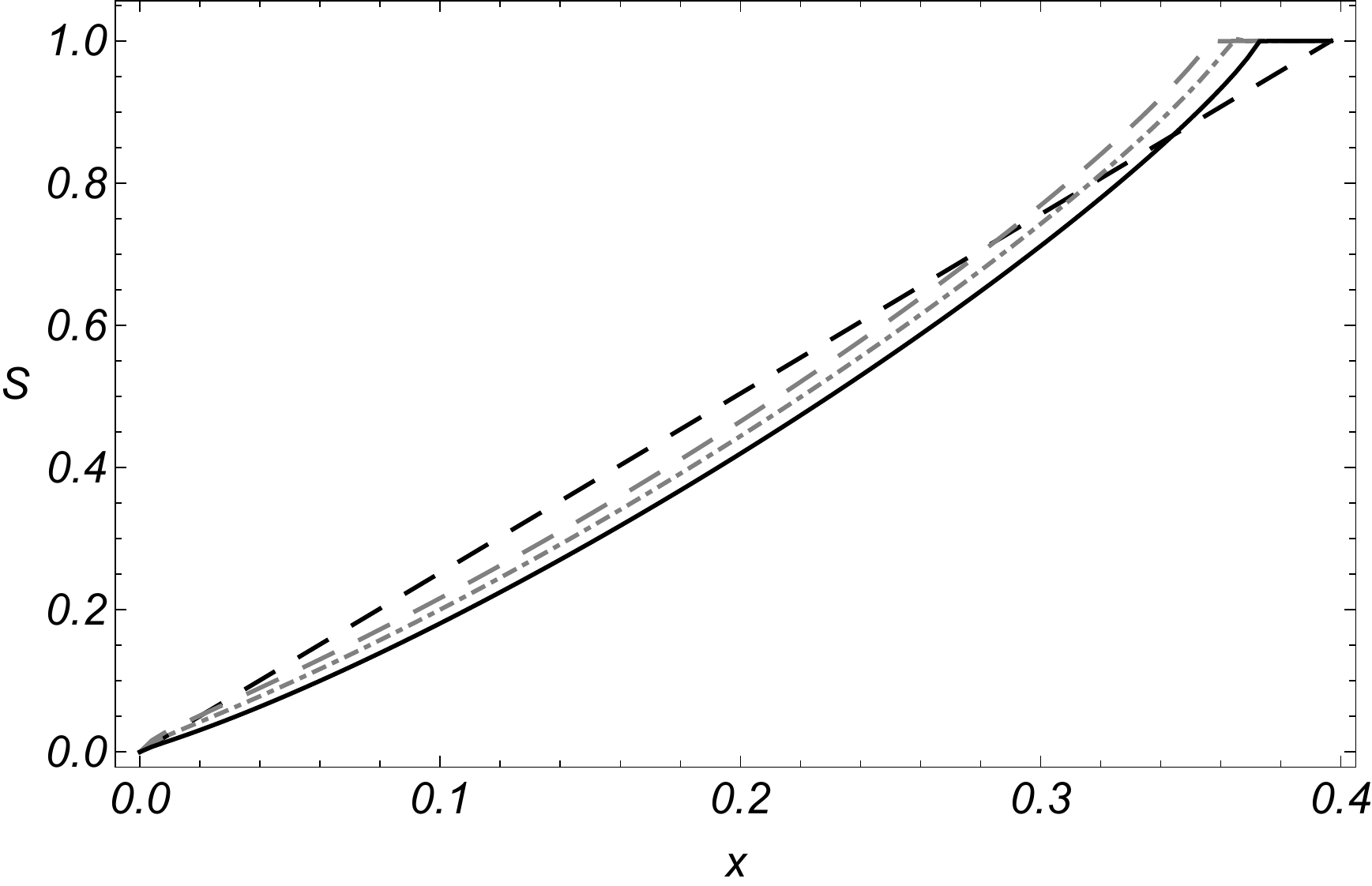}}}
}
\end{picture}  
\vskip 1. cm
\caption{Allen--Cahn dynamics with (D) and (N)--boundary conditions
and a linear initial profile (dashed black) connecting
the two phases. Two intermediate profiles (dashed gray) and the 
stationary profile (solid black) are depicted.
The domain of length $(b-a)=21/20\, \delta$ is discretized into $10^2$ finite elements; after $500$ 
dimensionless time steps, with $\Delta t=2.$ the distance between two subsequent profiles was $d^{(D)}=0.00013765$, 
and $d^{(N)}= 0.000108560687$ for the two cases.}
\label{f:Dbcs_L>delta}
\end{figure}
Assume a linear initial profile connecting the two phases and the length of the interval 
close to the length of the compacton, for instance $(b-a)=21/20 \, \delta$; the dynamics is
definitely similar to that in figures~\ref{f:Dbcs_L<delta} and \ref{f:Nbcs_L<delta}, where
the stationary profile is indeed formed by the compacton and the solution 
corresponding to the pure phase $S=1$, see figure~\ref{f:Dbcs_L>delta}.

\begin{figure}
\begin{picture}(100,120)
\put(-70,-30)
{
\resizebox{8cm}{!}{\rotatebox{0}{\includegraphics{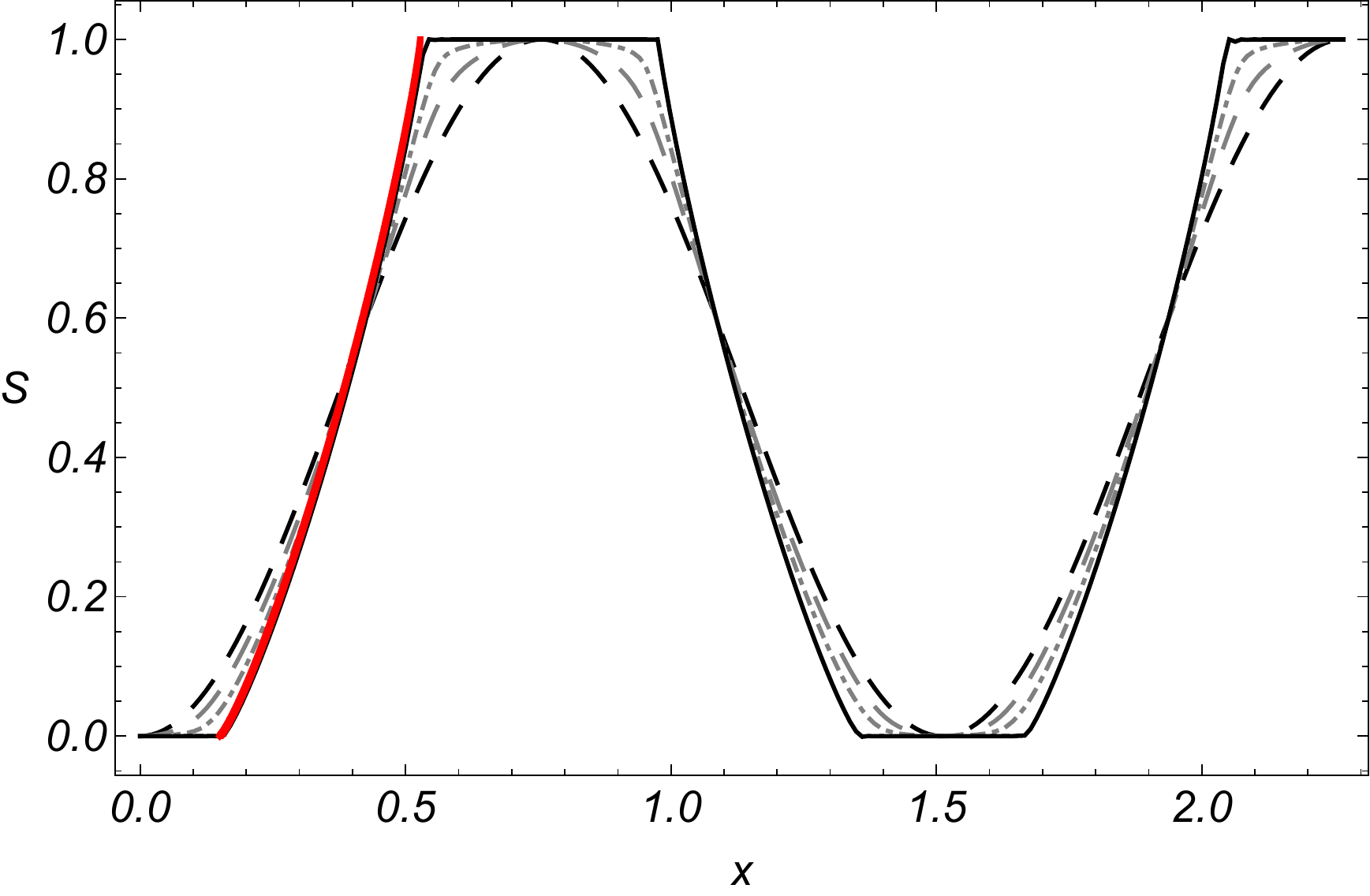}}}
}
\end{picture}  
\vskip 1. cm
\caption{Allen--Cahn dynamics with (D)--boundary conditions
and a co--sinusoidal initial profile (dashed black) connecting
the two phases. Two intermediate profiles (dashed gray), the 
stationary profile (solid black) and the compacton profile (solid red) are depicted. 
The domain of length $(b-a)=6\, \delta$ is discretized into $2\, 10^2$ finite elements; the 
time needed 
to get a distance of $10^{-3}$ between two subsequent profiles is $t_f=29$.}
\label{f:Dbcs_L>>delta}
\end{figure}

\begin{figure}
\begin{picture}(100,120)
\put(-70,-30)
{
\resizebox{8cm}{!}{\rotatebox{0}{\includegraphics{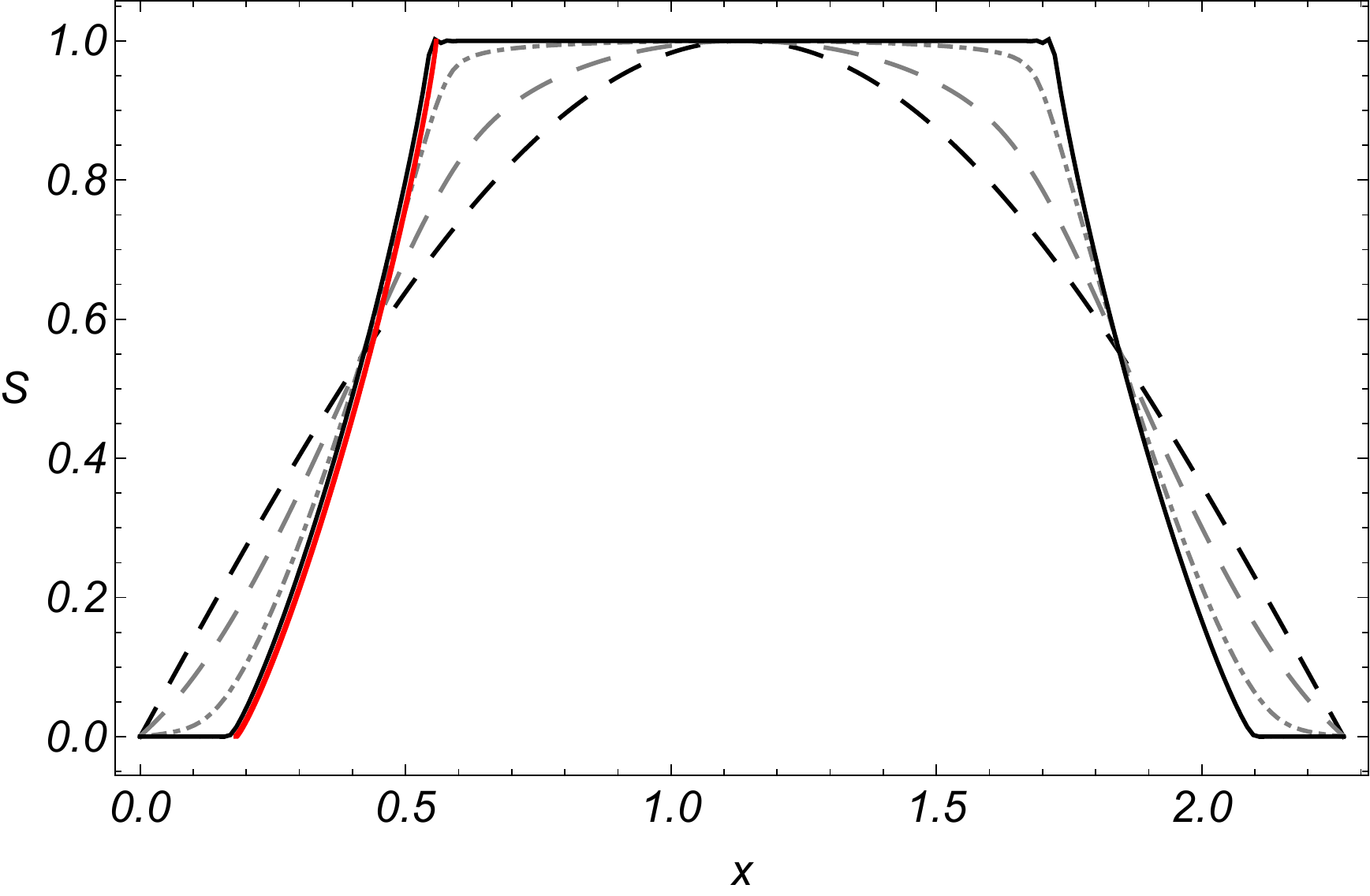}}}
}
\end{picture}  
\vskip 1. cm
\caption{Allen--Cahn dynamics with (N)--boundary conditions
and a one lobe sinusoidal initial profile (dashed black) connecting
the two phases. Two intermediate profiles (dashed gray), the 
stationary profile (solid black) and the compacton profile (solid red) are depicted. 
The domain of length $(b-a)=6\, \delta$ is discretized into $2\, 10^2$ finite elements; the 
time needed 
to get a distance of $10^{-3}$ between two subsequent profiles is $t_f=33$.}
\label{f:Nbcs_L>>delta}
\end{figure}

Consider now an interval whose length is $(b-a)=6\, \delta$, in this case, depending on the initial conditions,
one or more compactons can form in the domain so that oscillating solutions can indeed correspond to stationary 
states of the Allen--Cahn dissipative dynamics. In figures~\ref{f:Dbcs_L>>delta} and \ref{f:Nbcs_L>>delta} two distinct cases are exhibited which correspond to (D)
and (N)--boundary conditions.


\section{Conclusions}
\label{s:conclusioni}
We have considered a generalized Allen--Cahn equation
deduced from a Landau energy functional with a non--constant
higher order stiffness vanishing at the two 
pure phases. We have solved analytically the stationary problem
and deduced the existence of the so--called compactons.
We have also showed the possibilities of piecewise stationary solutions 
made of the superposition of compactons and constant pure phase
profiles. 

In a case of particular physical interest the compacton problem 
has been solved explicitly and the main physical features of 
such profiles connecting a liquid and a gas phase in a capillarity tube 
have been deduced. 

The dynamics has been studied numerically and the compacton formation 
has been described in detail. In this framework one of the most 
relevant result we discussed is the possibility that, due to the 
presence of compactons and by choosing properly the 
initial condition, the dissipative Allen--Cahn evolution 
can result in the formation of periodic profiles connecting the 
two pure phases. This stationary profiles pops up as the long time 
limit of the dynamical problem. It is important to stress that 
this possibility is ruled out in the standard Allen--Cahn dynamics.

\appendix

\section{Derivation of the Allen--Cahn equation}
\label{s:ac}
For completeness we sketch the derivation of the Allen--Cahn 
equation \eqref{ac010} in the case in which the higher--order stiffness coefficient 
is not constant. 

The gradient $\rr{grad}\,H$ of the Landau functional \eqref{grad-f} in the 
space $L^2(\Omega)$ is a function in such a space such that 
\begin{displaymath}
\frac{\rr{d}}{\rr{d}s}H(u+sv)\Big|_{s=0}
=
\int_\Omega v\,\rr{grad}\,H\,\rr{d}x
\end{displaymath}
for any $v\in L^2(\Omega)$.
In other words, 
the derivative of the function in any direction is equal to the 
scalar product of such a function with the one characterizing the 
direction. 
By \eqref{grad-f} it follows that 
\begin{equation}
\label{varp}
\begin{array}{l}
{\displaystyle 
\frac{\rr{d}}{\rr{d}s}H(u+sv)
=
\vphantom{\bigg\{_\}}
}
\\
{\displaystyle 
\phantom{mmm}
\int_\Omega
\Big[
     \frac{1}{2}\varepsilon'(u+sv)\|\nabla u+s\nabla v\|^2v
}
\\
{\displaystyle 
\phantom{mmmmm}
     +\varepsilon(u+sv)(\nabla u+s\nabla v)\cdot\nabla v
}
\\
{\displaystyle 
\phantom{mmmmm}
     +W'(u+sv)v
\Big]
\,
\rr{d}x
}
\end{array}
\end{equation}
Hence, 
\begin{displaymath}
\begin{array}{l}
{\displaystyle 
\frac{\rr{d}}{\rr{d}s}H(u+sv)\Big|_{s=0}
=
\vphantom{\bigg\{_\}}
}
\\
{\displaystyle 
\phantom{mm}
\int_\Omega
\Big[
     \frac{1}{2}\varepsilon'(u)\|\nabla u\|^2v
     +\varepsilon(u)\nabla u\cdot\nabla v
     +W'(u)v
\Big]
\,
\rr{d}x
}
\end{array}
\end{displaymath}
For $f,g,h:\bb{R}^3\to\bb{R}$ sufficiently regular, we 
recall the Green identity
\begin{equation}
\label{green}
\begin{array}{l}
{\displaystyle
\int_\Omega f \nabla g\cdot\nabla h\,\rr{d}x
\vphantom{\bigg\{_\}}
=
}
\\
{\displaystyle
\phantom{mmm}
-\int_\Omega h\nabla\cdot(f\nabla g)\,\rr{d}x
+\int_{\partial\Omega}hf\frac{\partial g}{\partial n}\,\rr{d}S
}
\end{array}
\end{equation}
with
$\partial\Omega$ the boundary of $\Omega$
and 
$\partial g/\partial n$ the derivative in the direction 
orthogonal to the boundary.
We then get
\begin{displaymath}
\begin{array}{l}
{\displaystyle
\frac{\rr{d}}{\rr{d}s}H(u+sv)\Big|_{s=0}
\vphantom{\bigg\{_\}}
}
\\
{\displaystyle
\phantom{mm}
=
\int_\Omega
\Big[
     \frac{1}{2}\varepsilon'(u)\|\nabla u\|^2
     -\nabla\cdot(\varepsilon(u)\nabla u) 
     +W'(u)
\Big]
v
\,
\rr{d}x
\vphantom{\bigg\{_\big\}}
}
\\
{\displaystyle
\phantom{mm=}
 +\int_{\partial\Omega}v\varepsilon(u)\frac{\partial u}{\partial n}\,\rr{d}S
}
\end{array}
\end{displaymath}
Moreover, recalling the properties of the divergence operator we get 
\begin{equation}
\label{var010}
\begin{array}{l}
{\displaystyle
\frac{\rr{d}}{\rr{d}s}H(u+sv)\Big|_{s=0}
\vphantom{\bigg\{_\}}
}
\\
{\displaystyle
\phantom{i}
=
\int_\Omega
\Big[
     -\frac{1}{2}\varepsilon'(u)\|\nabla u\|^2
     -\varepsilon(u)\Delta u
     +W'(u)
\Big]
v
\,
\rr{d}x
\vphantom{\bigg\{_\big\}}
}
\\
{\displaystyle
\phantom{i=}
 +\int_{\partial\Omega}v\varepsilon(u)\frac{\partial u}{\partial n}\,\rr{d}S
}
\end{array}
\end{equation}
Finally, 
from this equality, in the Lebesgue space of functions such that 
the normal derivative to the boundary of $\Omega$ vanishes, we have that 
\begin{displaymath}
\rr{grad}\,H(u)
=
     -\frac{1}{2}\varepsilon'(u)\|\nabla u\|^2
     -\varepsilon(u)\Delta u 
     +W'(u)
\end{displaymath}
which yields the Allen--Cahn equation \eqref{ac010}.

\section{Integral computations}
\label{s:integrali}
The integrals \eqref{inter27} and \eqref{inter32} can 
be computed by using the properties of the gamma and beta functions. 

Recall the definition of the beta function and 
that of the incomplete beta function
\begin{equation}
\label{app00}
B(p,q)
=
\int_0^1t^{p-1}(1-t)^{q-1}\,\rr{d}t
\end{equation}
and
\begin{equation}
\label{app10}
B(x,p,q)
=
\int_0^xt^{p-1}(1-t)^{q-1}\,\rr{d}t
\end{equation}
with $\textrm{Re}(p),\textrm{Re}(q)>0$.
It is immediate to prove that 
\begin{equation}
\label{app12}
B(p,q)=B(1,p,q)
\end{equation}
and 
\begin{equation}
\label{app14}
\frac{\rr{d}}{\rr{d}x}
 [xB(x,p,q)-B(x,p+1,q)]
=
B(x,p,q)
\end{equation}

In the following we shall also need some properties of the 
gamma function. 
Recall its definition 
\begin{equation}
\label{app20}
\Gamma(p)
=
\int_0^\infty t^{p-1}\,e^{-t}\,\rr{d}t
\end{equation}
with $\textrm{Re}(p)>0$, and the two properties
\begin{equation}
\label{app30}
\Gamma(p+1)=p\Gamma(p)
\;\textrm{ and }\;
\Gamma(1-p)\Gamma(p)=\frac{\pi}{\sin(\pi p)}
\end{equation}

The beta function is related to the gamma function 
by the equality
\begin{equation}
\label{app40}
B(p,q)=
\frac{\Gamma(p)\Gamma(q)}{\Gamma(p+q)}
\end{equation}

Let $a$ be a real such that $0<a<1$, it is immediate 
to remark that 
\begin{equation}
\label{app50}
\int_0^x t^{-a}(1-t)^{a}\,\rr{d}t
=
B(x,1-a,1+a)
\end{equation}
Indeed, it is sufficient to let $p=1-a$ and $q=1+a$ and 
recall \eqref{app10}.

Moreover, 
\begin{displaymath}
\int_0^1 t^{-a}(1-t)^{a}\,\rr{d}t
=
B(1,1-a,1+a)
=B(1-a,1+a)
\end{displaymath}
where we used \eqref{app12}. On the other hand, 
by \eqref{app40} and the fact that $\Gamma(2)=1$, we have that 
\begin{displaymath}
B(1-a,1+a)
=
\Gamma(1-a)\Gamma(1+a)
=
\Gamma(1-a)a\Gamma(a)
\end{displaymath}
where in the last step we have used the first of 
\eqref{app30}.
Hence, recalling the second of \eqref{app30}, we have that 
\begin{equation}
\label{app60}
\int_0^1 t^{-a}(1-t)^{a}\,\rr{d}t
=
\frac{\pi a}{\sin(\pi a)}
\end{equation}

Finally, with simple algebra, we get that 
\begin{displaymath}
\begin{array}{l}
{\displaystyle
 \int_0^1\rr{d}x
 \int_0^x\rr{d}t\, t^{-a}(1-t)^{a}
 \vphantom{\bigg\{_\}}
}
\\
{\displaystyle
 \phantom{mm}
 =B(1-a,1+a)-
 \int_0^1\rr{d}x\, B(x,1-a,1+a)
}
\\
\end{array}
\end{displaymath}
By \eqref{app14} we find 
\begin{displaymath}
 \int_0^1\rr{d}x
 \int_0^x\rr{d}t\, t^{-a}(1-t)^{a}
=
B(2-a,1+1)
\end{displaymath}
On the other hand, by using the properties of the gamma 
and the beta functions as above and recalling that $\Gamma(3)=2$, 
we have that
\begin{displaymath}
\begin{array}{rcl}
B(2-a,1+1)
&\!\!=&\!\!
{\displaystyle
 \frac{\Gamma(2-a)\Gamma(1+a)}{\Gamma(3)}
 =
 \frac{\Gamma(2-a)\Gamma(1+a)}{2}
 \vphantom{\bigg\{_\}}
}\\
&\!\!=&\!\!
{\displaystyle
 \frac{1}{2}(1-a)a\Gamma(1-a)\Gamma(a)
}\\
{\displaystyle
 \vphantom{\bigg\{_\}}
}\\
\end{array}
\end{displaymath}
By the second of \eqref{app30} we thus get 
\begin{equation}
\label{app70}
 \int_0^1\rr{d}x
 \int_0^x\rr{d}t\, t^{-a}(1-t)^{a}
=
 \frac{1}{2}(1-a)a
 \frac{\pi}{\sin(\pi a)}
\end{equation}

\begin{acknowledgments}
We wish to express our thanks to R.\ Benzi and P.\ Butt\`a for very useful 
discussions. 
\end{acknowledgments}

\end{document}